 \documentclass[final,5p,times,twocolumn,authoryear]{elsarticle}

\usepackage{amssymb}
\usepackage{lipsum}

\usepackage{hyperref}
\usepackage{academicons}
\usepackage{xcolor}[dvipsnames]
\usepackage{fontawesome5}

\definecolor{orcidlogocol}{rgb}{0.65, 0.807, 0.223}

\newcommand{\orcid}[1]{$\,$\href{https://orcid.org/#1}{\textcolor{orcidlogocol}{\faOrcid}}}

\journal{Nuclear Physics B}

\begin{document}

\begin{frontmatter}



\title{Primordial black holes as a dark matter candidate - a brief overview}


\author[first]{Anne M. Green\orcid{0000-0002-7135-1671}}
\affiliation[first]{organization={University of Nottingham},
            addressline={School of Physics and Astronomy}, 
            city={Nottingham},
            postcode={NG7 2RD}, 
            state={},
            country={UK}}

\begin{abstract}
Historically the most popular dark matter candidates have been new elementary particles, such as Weakly Interacting Massive Particles and axions. However Primordial Black Holes (PBHs), black holes formed from overdensities in the early Universe, are another possibility. The discovery of gravitational waves from mergers of tens of Solar mass black hole binaries by LIGO-Virgo has generated a surge in interest in PBH dark matter. We overview the formation of PBHs, observational probes of their abundance, and some of the key open questions in the field.
\end{abstract}




\end{frontmatter}




\section{Motivation and history}
\label{sec:intro}

Primordial Black Holes (PBHs) fulfill all of the necessary requirements to be a good dark matter (DM) candidate. They are cold, non-baryonic and stable, and can be formed in the right abundance to be the DM. \citet{1967SvA....10..602Z} and \citet{1971MNRAS.152...75H} found that Primordial Black Holes may form from overdensities in the early Universe. As they form before nucleosynthesis, PBHs are non-baryonic. PBHs evaporate by emitting Hawking radiation \citep{Hawking:1975vcx}, however a PBH is cosmologically stable (its lifetime is longer than the age of the Universe) if its initial mass is greater than $\sim 10^{15} \, {\rm g}$ \citep{Hawking:1975vcx,Page:1976df}. Unlike most other DM candidates (for instance Weakly Interacting Massive Particles, axions, sterile neutrinos,...) PBHs aren't a new particle. However, as we will see in Sec.~\ref{sec:form}, their formation does typically require `Beyond the Standard Model' physics, e.g.~inflation.

It was realised that PBHs are a DM candidate already in the 1970s \citep{1971MNRAS.152...75H,Chapline:1975ojl}. A wave of interest
in Solar mass PBH DM occurred in the 1990s, due to the excess of microlensing events observed in the MACHO collaboration's 2 year data set \citep{MACHO:1996qam}. The number of events was larger than expected from known stellar populations, and consistent with roughly half of the Milky Way halo being in Solar mass compact objects. \citet{Nakamura:1997sm} pointed out that if a significant fraction of the DM is in the form of $\sim$ Solar mass PBHs, then PBH binaries would form in the early Universe and (if those binaries survive to the present day) gravitational waves from their coalescence would be detectable by LIGO.

In 2016 LIGO-Virgo announced the discovery of gravitational waves from mergers of tens of Solar mass black holes (BHs) \citep{LIGOScientific:2016aoc}. Shortly afterwards several papers appeared \citep{Bird:2016dcv,Clesse:2016vqa,Sasaki:2016jop}, suggesting that (some of) these BHs could be primordial, rather than astrophysical, and that these PBHs could also be a significant component of the DM.
This has led to another, much larger, wave of interest in PBH DM, with the number of papers written on PBHs rising to $\sim 300$ per year.

In Sec.~\ref{sec:form} we describe the formation of PBHs, focusing on the collapse of large density perturbations during radiation domination. In Sec.~\ref{sec:obs} we overview observational probes of the abundance of PBHs. Finally, in Sec.~\ref{sec:openq}, we review some of the key open questions regarding PBH dark matter, before concluding with a summary in Sec.~\ref{sec:summary}. 

This brief overview is aimed at readers with knowledge of DM in general, but not PBHs specifically. For longer, more detailed reviews of PBHs as a dark matter candidate with extensive reference lists, see \citet{Carr:2020xqk} and \citet{Green:2020jor}. For a `positivist perspective' on observational evidence for PBHs see \citet{Carr:2023tpt}, and for future observational searches for PBHs see the Snowmass 2021 white paper \citep{Bird:2022wvk}.

\section{Formation}
\label{sec:form}

We will focus on the most popular, and arguably minimal, PBH formation mechanism, namely the collapse of large density perturbations during radiation domination \citep{1967SvA....10..602Z,1971MNRAS.152...75H,Carr:1974nx}. These are the same density perturbations that on large scales form galaxies and large scale structure, but on shorter length scales and (if a non-negligible number of PBHs are to form) with a much larger amplitude. If a region is sufficiently overdense then when it enters the horizon (i.e.~becomes comparable in length scale to the observable Universe at that time), gravity rapidly overcomes pressure forces and it collapses to form a BH. Other early Universe processes which produce overdensities can also lead to PBH formation. Examples include the collapse of cosmic string loops \citep{Hawking:1987bn,Polnarev:1988dh}, collisions of bubbles from phase transitions \citep{Hawking:1982ga}, fragmentation of an inflaton scalar condensate \citep{Cotner:2016cvr} and the collapse of density perturbations during an early period of matter domination \citep{Khlopov:1980mg}.

In Sec.~\ref{subsec:cldp} we review the essentials of the formation of PBHs from the collapse of large density perturbations during radiation domination, including the criterion for PBH formation and the PBH mass and abundance. In Sec.~\ref{subsec:inflation} we overview the generation of large density perturbations by inflation. Finally in Sec.~\ref{subsec:refinements} we discuss refinements to the calculations of the PBH abundance and mass function.

\subsection{Collapse of large density perturbations}
\label{subsec:cldp}

The criterion for PBH formation during radiation domination, and the PBH mass and abundance were calculated by \citet{Carr:1975qj}. The criterion for PBH formation is most easily specified in terms of the density contrast, at horizon crossing,  $\delta \equiv (\rho- \bar{\rho})/\bar{\rho}$, where $\rho$ is the density of the region and ${\bar{\rho}}$ is the average background density. A PBH will form if the density contrast exceeds a critical value, $\delta_{\rm c}$, which is of order the equation of state parameter $w \equiv p/\rho = 1/3$~\footnote{Here and throughout we use natural units with $c=1$.}. The mass of the PBH formed, $M_{\rm PBH}$, is roughly equal to the mass within the horizon, $M_{\rm H}$, at that time, $t$,:
\begin{equation}
\label{eq:mpbh}
M_{\rm PBH} \approx M_{\rm H} \sim 10^{15} \, {\rm g} \, \left( \frac{t}{10^{-23} \, {\rm s}}
\right) \sim M_{\odot} \, \left( \frac{t}{10^{-6} \, {\rm s}}
\right) \,.
\end{equation}

\begin{figure}
	\centering 
	\includegraphics[width=0.49\textwidth, angle=0]{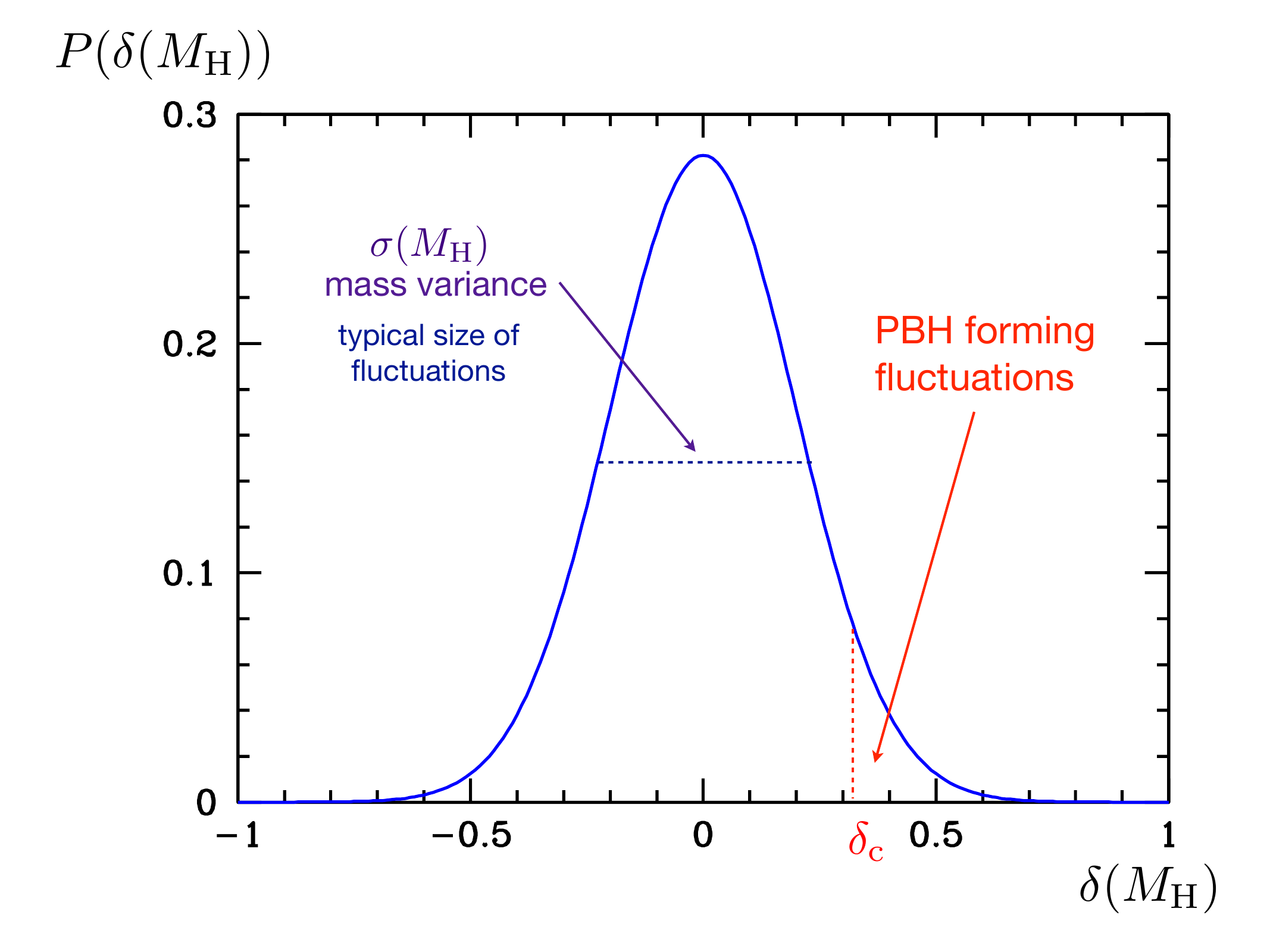}	
	\caption{An illustration of the probability distribution, $P(\delta(M_{\rm H}))$, of the density contrast, $\delta(M_{\rm H})$. PBHs form from rare large fluctuations with $\delta(M_{\rm H})$ greater than the critical value, $\delta_{\rm c}$, which is of order $1/3$.} 
	\label{fig_prob}%
\end{figure}

The initial PBH mass fraction, $\beta$, is defined as the fraction of the Universe in regions which are over-dense enough to form a PBH. It can be calculated by integrating the probability distribution of the density contrast on the scale of interest (specified in terms of the horizon mass, $M_{\rm H}$), $P(\delta(M_{\rm H}))$, above the critical value $\delta_{\rm c}$:
\begin{equation}
\beta(M_{\rm H}) \sim \int_{\delta_{\rm c}}^{\infty} P(\delta(M_{\rm H})) \, {\rm d} \delta(M_{\rm H}) \,.
\end{equation}
Assuming (for now) that the probability distribution of the density contrast is gaussian then
\begin{equation}
\label{eq:beta1}
\beta(M_{\rm H}) \sim {\rm erfc} \left( \frac{\delta_{\rm c}}{\sqrt{2} 
\sigma(M_{\rm H})} \right) \,,
\end{equation}
where {\rm erfc} is the complementary error function and $\sigma(M_{\rm H})$, the mass variance, is the typical size of the fluctuations. See Fig.~\ref{fig_prob} for a visual illustration. As we will see shortly, $\beta$ has to be small and hence $\sigma(M_{\rm H}) \ll \delta_{\rm c}$ so that
\begin{equation}
\label{eq:beta2}
\beta(M_{\rm H}) \sim \sigma(M_{\rm H}) \exp{ \left( - \frac{\delta_{\rm c}^2}{2 \sigma^2(M_{\rm H})} \right) } \,,
\end{equation}
and the PBH abundance depends exponentially on the typical size of the fluctuations, $\sigma(M_{\rm H})$.

\begin{figure}
	\centering 
	\includegraphics[width=0.49\textwidth, angle=0]{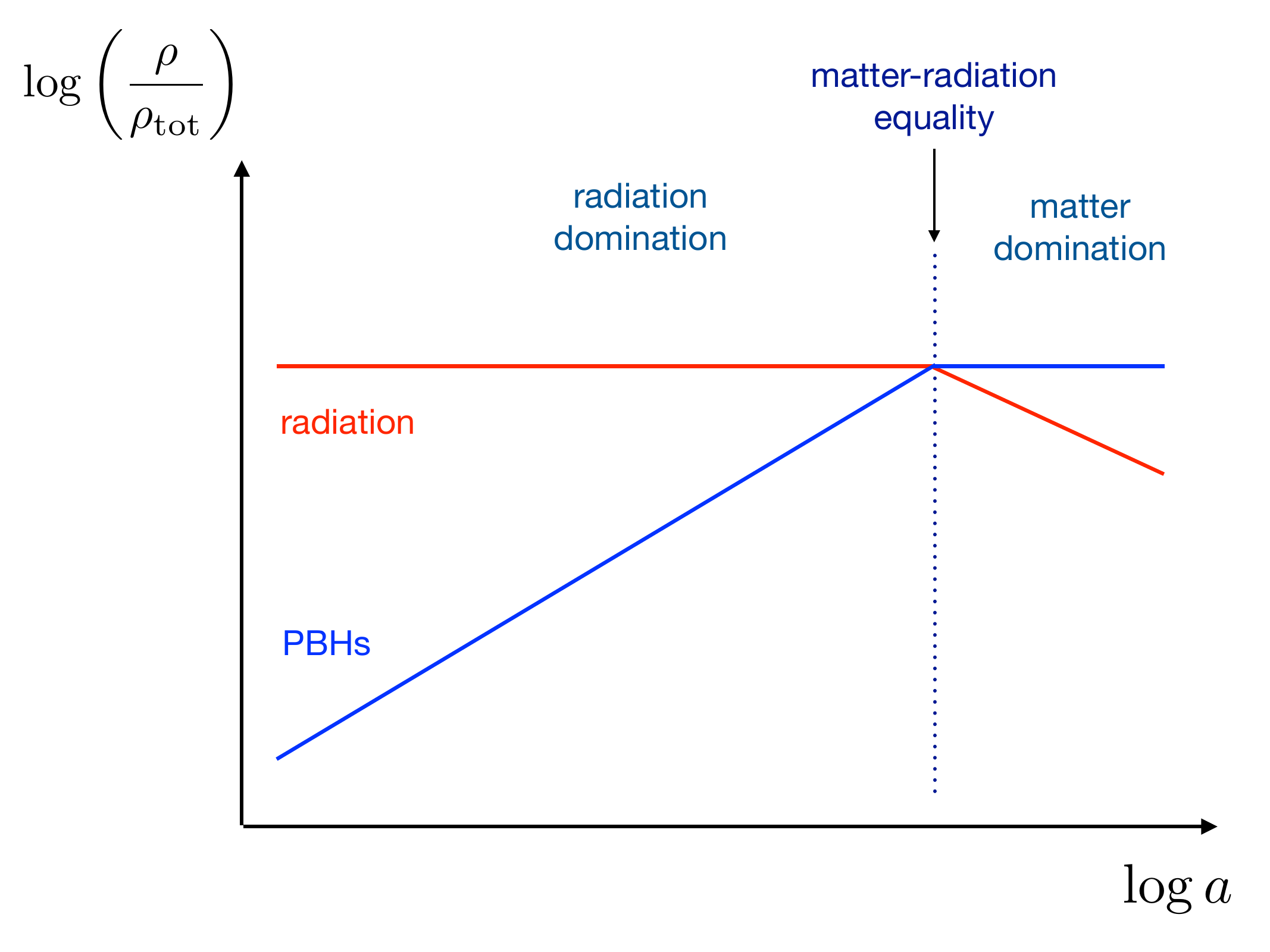}	
	\caption{An illustration of the evolution of the fraction of the total density in the form of radiation (in red) and PBHs (blue) as a function of the scale factor $a$. The dashed vertical line indicates the epoch of matter-radiation equality, where the Universe transitions from radiation to matter dominated.} 
	\label{fig_rho}%
\end{figure}

Next we need to relate the initial PBH mass fraction, $\beta$, to the fraction of the DM today in the form of PBHs, $f_{\rm PBH}=\rho_{\rm PBH}/\rho_{\rm DM}$, where $\rho_{\rm PBH}$ and $\rho_{\rm DM}$ are the PBH and DM energy densities. Since PBHs are matter their energy density dilutes as $\rho_{\rm PBH} \propto a^{-3}$, where $a$ is the scale factor that parameterises the expansion of the Universe. The radiation energy density dilutes as $\rho_{\rm rad} \propto a^{-4}$. Therefore during radiation domination the fraction of the total energy density in the form of PBHs grows:
\begin{equation}
\frac{\rho_{\rm PBH}}{\rho_{\rm rad}} \propto \frac{a^{-3}}{a^{-4}} \propto a \,.
\end{equation}
Consequently, as shown in Fig.~\ref{fig_rho}, to not exceed the observed DM abundance today, the initial PBH abundance has to be very small, specifically
\begin{equation}
\beta(M_{\rm H}) \sim 10^{-9} f_{\rm PBH} \left( \frac{M}{M_{\rm PBH}} \right) \,.
\end{equation}
Lighter PBHs form earlier and therefore have to have a smaller initial abundance, $\beta$, to make up a fixed fraction of the DM today, $f_{\rm PBH}$.

On cosmological scales the amplitude of the primordial perturbations is measured, by Planck observations of the Cosmic Microwave Background (CMB), to be $\sigma \approx 10^{-5}$~\citep{Planck:2018jri}. If the primordial perturbations are very close to scale-invariant (i.e.~have a similar amplitude on all scales) then the number of PBHs formed will be completely negligible. Using Eqs.~(\ref{eq:beta1}-\ref{eq:beta2})
\begin{equation}
\beta(M_{\rm H}) \sim {\rm erfc}(10^{5}) \sim \exp{(-10^{10})} \,.
\end{equation}

\begin{figure}
	\centering 
	\includegraphics[width=0.45\textwidth, angle=0]{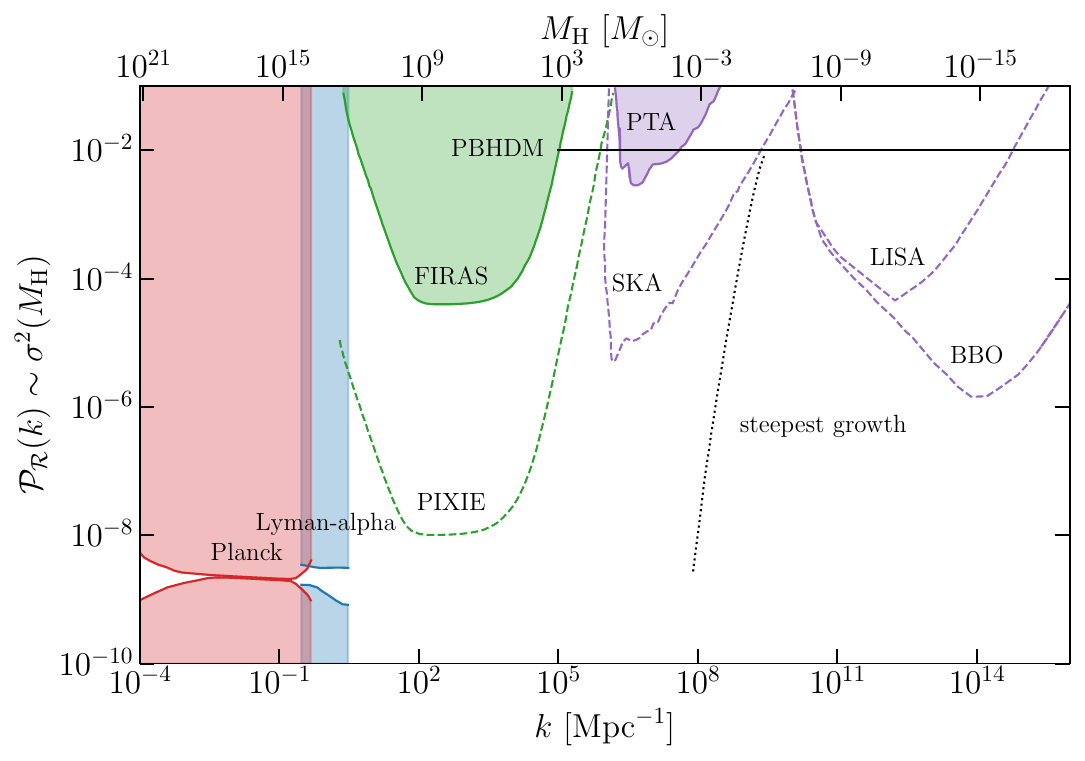}	
	\caption{Constraints on the primordial power spectrum of the curvature perturbation, ${\cal P}_{\cal R}(k)$, which is of order the square of the mass variance, $\sigma^2(M_{\rm H})$, as a function of both horizon mass, $M_{\rm H}$, in Solar masses, $M_{\odot}$, (top horizontal axis) and comoving wavenumber, $k$, in ${\rm Mpc}^{-1}$ (bottom horizontal axis).
  The constraints shown are (from left to right) from the CMB temperature angular power spectrum~\citep{Planck:2018jri} (in red), the Lyman-$\alpha$ forest \citep{Bird:2010mp} (blue), CMB spectral distortions~\citep{Fixsen:1996nj} (green) and pulsar timing array limits on gravitational waves~\citep{Byrnes:2018txb} (magenta). The sensitivity of a future PIXIE-like spectral distortion experiment \citep{Chluba:2019nxa} (blue) and gravitational waves searches from SKA, LISA and BBO (magenta) \citep{Inomata:2018epa,Chluba:2019nxa} are shown as dotted lines. In each case the currently excluded regions are shaded. The horizontal black line shows the approximate amplitude, ${\cal P}_{\cal R}(k) \sim 10^{-2}$, required to form an interesting number of PBHs. The dotted black line shows the steepest growth expected in single-field inflation models, $\sim k^{4}$, \citep{Byrnes:2018txb}. This figure was created using the PBHbounds code \citep{PBHbounds}, similar figures appear in \citet{Byrnes:2018txb,Inomata:2018epa,Chluba:2019nxa}.} 
	\label{fig_PSconstraints}%
\end{figure}

To form an interesting number of PBHs the primordial perturbations must be significantly larger, $\sigma \sim 0.1$, on some small scale than they are on cosmological scales.   
Figure \ref{fig_PSconstraints} shows the constraints on the primordial power spectrum of the curvature perturbation, ${\cal P}_{\cal R}(k)$, which is of order the mass variance squared: ${\cal P}_{\cal R}(k) \sim \sigma^2(M_{\rm H})$. On cosmological scales the amplitude is accurately measured and small. On smaller scales there are upper limits from CMB spectral distortions~\citep{Carr:1993aq} and 
scalar induced gravitational waves (SIGWs), tensor perturbations generated at second order by large scalar, i.e.~density, perturbations \citep{Ananda:2006af}\footnote{Recently pulsar timing arrays, including NANOGrav, have detected a background of low frequency gravitational waves which could be explained by inspiralling supermassive BHs, or by via various cosmological processes, including SIGWs from scalar perturbations with amplitude ${\cal P}_{\rm R}(M_{\rm H} \sim 10^{-3} M_{\odot}) ~\sim 0.1$~\citep{NANOGrav:2023hvm}.}. Greater sensitivity may be obtained in future, from gravitational wave searches by e.g.~SKA, LISA and BBO \citep{Inomata:2018epa,Chluba:2019nxa} or a PIXIE-like spectral distortion experiment \citep{Chluba:2019nxa}.

\subsection{Producing large inflationary density perturbations}
\label{subsec:inflation}

The primordial perturbations are thought to be produced by inflation, a period of accelerated expansion in the early Universe, driven by a scalar field, $\phi$, `slow-rolling'
along its potential. In slow-roll inflation the mass variance is given by
\begin{equation}
\label{eq:sigma-sr}
\sigma \propto \frac{V^{3/2}}{V^{\prime}} \,,
\end{equation}
where $V$ is the potential and $V^{\prime}$ is its derivative with respect to $\phi$. It was pointed out by \citet{Ivanov:1994pa} that a plateau in the potential (with small $V^{\prime}$) can generate large perturbations, that form an interesting number of PBHs. However in the ultra-slow-roll limit, $V^{\prime} \rightarrow 0$, the slow-roll approximation breaks down, and a numerical calculation is required to accurately calculate the amplitude of the perturbations.

A successful PBH producing inflation model has to fulfill three requirements:
\begin{enumerate}
\item match the measured amplitude and scale dependence of the perturbations on cosmological scales,
\item have significantly larger perturbations, $\sigma(M_{\rm H}) \sim 0.1$, on some smaller scale,
\item inflation ends.
\end{enumerate}
A single-field inflation model typically has to be fine-tuned to satisfy all three of these requirements. An example is models with a very small local minimum in the potential. In this case the field traverses this feature with reduced speed, generating large perturbations, and then reaches a region where the potential is steep and inflation ends \citep{Ballesteros:2017fsr,Hertzberg:2017dkh}. The steepest growth in the power spectrum that can be produced in single-field inflation models is of order $\sim k^4$~\citep{Byrnes:2018txb}.

These three requirements can be more easily satisfied in multi-field inflation models, where for instance a different field is responsible for generating the perturbations than for ending or driving inflation. A commonly studied possibility is hybrid inflation, where inflation ends due to a second field undergoing a `waterfall' phase transition and large perturbations can be generated at the phase transition~\citep{Garcia-Bellido:1996mdl}. If the waterfall transition is `mild' (occurring neither too quickly nor too slowly) the power spectrum can be consistent with observations on cosmological scales and also have a broad, PBH forming, peak on smaller scales~\citep{Clesse:2015wea}. Other possibilities include running mass inflation, double inflation, an axion-like curvaton, a reduction in the sound speed and multi-field models with rapid turns in field space. For detailed reviews see \citet{Ozsoy:2023ryl} and \citet{Escriva:2022duf}.

\subsection{Refinements to calculation}
\label{subsec:refinements}

To accurately compare theoretical predictions with observations (and reliably ascertain whether or not PBHs make up a given fraction of the DM) accurate calculations of the PBH abundance, and mass function, are required. We will therefore conclude this section, by discussing refinements to the calculations discussed in Sec.~\ref{subsec:cldp} above.

The threshold for PBH formation in fact depends on the shape of the perturbation~\citep{Harada:2013epa,Musco:2018rwt}, which depends on the primordial power spectrum~\citep{Germani:2018jgr}, and is best specified in terms of the compaction function~\citep{Musco:2018rwt,Germani:2018jgr,Escriva:2019phb}. For a detailed review see \citet{Escriva:2022duf}.

At phase transitions the pressure is reduced and consequently the threshold for PBH formation, $\delta_{\rm c}$, is reduced. Therefore the abundance of PBHs formed, for a fixed mass variance, is increased. In particular the horizon mass at the QCD phase transition is of order a Solar mass, leading to enhanced formation of PBHs of this mass~\citep{Jedamzik:1996mr}. If the primordial perturbations have a large constant amplitude, $\sigma(M_{\rm H}) \sim 0.1$, over a wide range of scales, then the formation of PBHs with $M_{\rm PBH} \sim 10^{-5}, 1$ and $10^{7} M_{\odot}$ will be enhanced due to the electroweak phase transition, the QCD phase transition and $e^{+} e^{-}$ annihilation respectively~\citep{Carr:2023tpt}.

\citet{Niemeyer:1997mt} pointed out that, due to critical phenomena in gravitational collapse, the mass of a PBH depends on the size of the fluctuation from which it forms: $M_{\rm PBH} = k M_{\rm H} (\delta -\delta_{\rm c})^{\gamma}$ where $k$ and $\gamma$ are constants. This relationship has been verified using numerical simulations, e.g.~\citet{Musco:2008hv}. Consequently even if PBHs all form at the same time, there is a spread in their masses i.e.~the PBH mass function is not expected to be a delta function.

It was pointed out in the 1990s that since PBHs form from rare large fluctuations, their abundance depends sensitively on the shape of the tail of the density contrast probability distribution~\citep{Bullock:1996at,Ivanov:1997ia}. More recently it has been realised that even if the curvature perturbations are gaussian, the distribution of large density perturbations won't be, because the relationship between curvature perturbations and density perturbation is non-linear~\citep{Franciolini:2018vbk,DeLuca:2019qsy,Young:2019yug}. Furthermore, in ultra-slow-roll inflation (which, as discussed above, is required in single-field models to produce large perturbations) the probability distribution of the curvature perturbations may not be gaussian. See Sec.~\ref{subsec:USR} for further discussion.

\section{Observational constraints/signatures}
\label{sec:obs}

In this Section we focus on observational constraints on the present day abundance of PBHs. In Sec.~\ref{subsec:currentconstraints} we briefly review current constraints, calculated assuming that the PBHs have a delta-function mass function (DF MF). In Sec.~\ref{subsec:futureconstraints} we overview potential improvements to these constraints in future, and in Sec.~\ref{subsec:emf} we discuss constraints on more realistic extended mass functions.
We focus here on constraints on the fraction of the DM in the form of PBHs. \citet{Carr:2023tpt} discusses a wide range of observations which could potentially be explained by PBHs (with $f_{\rm PBH} \leq 1)$.

\subsection{Current constraints}
\label{subsec:currentconstraints}
\begin{figure}
	\centering 
	\includegraphics[width=0.45\textwidth, angle=0]{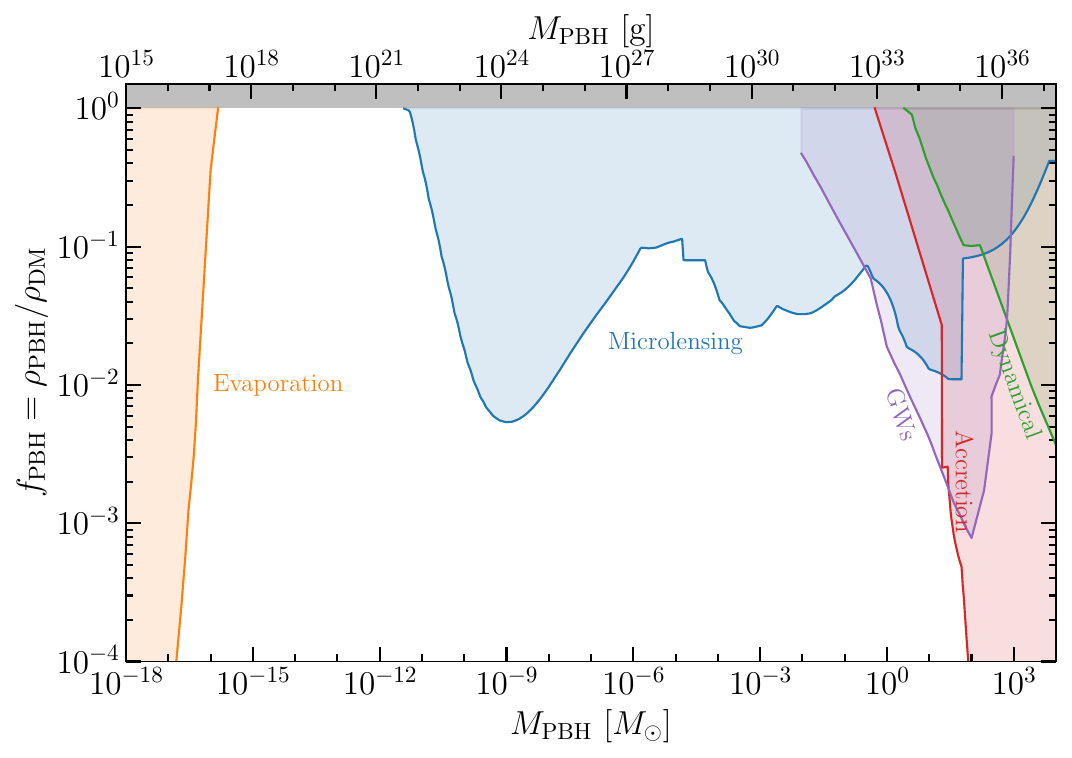}	
	\caption{Constraints on the fraction of DM in the form of PBHs, $f_{\rm PBHs}$, as a function of mass, $M_{\rm PBH}$, assuming all PBHs have the same mass. 
   The bounds shown are (from left to right) from evaporation (in orange), microlensing (blue), gravitational waves (purple), accretion (red) and dynamical (green).
   For each type of bound the tightest constraint at each mass is shown and the shaded regions are excluded. Figure created using the PBHbounds code~\citep{PBHbounds}.} 
	\label{fig_constraints}%
\end{figure}

Figure~\ref{fig_constraints} shows the current constraints\footnote{\citet{Boehm:2020jwd} argue that PBHs in the early Universe shouldn’t be described by the Schwarzschild metric, and point out that with an alternative, Thakurta, metric the mass of a PBH grows and hence the evaporation and gravitational wave constraints change significantly.} on the fraction of DM in the form of PBHs, $f_{\rm PBH}=\rho_{\rm PBH}/\rho_{\rm DM}$, as a function of PBH mass, $M_{\rm PBH}$, assuming all PBHs have the same mass, i.e.~that the PBH mass function is a delta function.
PBHs with initial mass $10^{15} \, {\rm g} \lesssim M_{\rm PBH} \lesssim 10^{17} \, {\rm g}$ are evaporating at a significant rate and their abundance is constrained by limits on the products of their evaporation, e.g.~MeV gamma-rays. Planetary and Solar mass PBHs are constrained by various microlensing observations. If all of the DM were in the form of multi-Solar mass PBHs there would in fact be several orders of magnitude more merger events than observed by LIGO-Virgo, so these observations constrain $f_{\rm PBH}$.
Multi-Solar mass PBHs can also be constrained by the consequences of their accretion, specifically the effect of the subsequent radiation on the recombination history of the Universe and hence the CMB, and also present day X-ray or radio emission. Massive PBHs are constrained via their dynamical effects on stars, in wide binaries and dwarf galaxies. For more detailed discussion of these constraints, including references to the original papers, see \citet{Carr:2020xqk} and \citet{Green:2020jor}.

As we see in Fig.~\ref{fig_constraints} it currently appears that planetary, Solar or multi-Solar mass PBHs making up all of the DM is excluded. There are uncertainties in these constraints, and some of the constraints have been revisited in recent years and significantly revised. However it seems unlikely (to the present author at least) that there are errors or uncertainties in multiple constraints such that $f_{\rm PBH}=1$ in this mass range is in fact allowed.
There is a range of masses, $10^{17} \, {\rm g} \lesssim M_{\rm PBH} \lesssim 10^{22} \, {\rm g}$, usually described as the `asteroid mass window', where PBHs can make up all of the DM, i.e.~$f_{\rm PBH} =1 $ is allowed. These light PBHs are hard to detect and we will discuss ideas and prospects in Sec.~\ref{subsec:asteroid}.

\subsection{Future constraints}
\label{subsec:futureconstraints}

The Snowmass 2021 Cosmic Frontier White Paper: Primordial Black Hole dark matter~\citep{Bird:2022wvk} discusses potential improvements to current constraints. Future MeV gamma-ray telescopes will tighten the evaporation constraints (see e.g. \citet{Coogan:2021rez} and Fig.~\ref{fig_asteroid}). However the rate at which PBHs evaporate decreases rapidly with increasing mass, ${\rm d} M_{\rm PBH}/{\rm d} t \sim M_{\rm PBH}^{-2}$ \citep{Hawking:1975vcx}, so the mass for which $f_{\rm PBH}=1$ is excluded (for a DF MF) by evaporation constraints will only increase by a factor of a few. Similarly it is hard to decrease the smallest mass for which $f_{\rm PBH}=1$ is excluded (for a DF MF) by microlensing observations. For $M_{\rm PBH} \lesssim 10^{-12} M_{\odot}$
the microlensing amplification is reduced due to finite source effects and wave optics (since the wavelength of light is similar to the Schwarzschild radius of the PBH), see \citet{Sugiyama:2019dgt} and references therein. A future microlensing survey of white dwarfs in the Large Magellanic Cloud could place much tighter constraints on PBHs with $M_{\rm PBH} \sim 10^{22} \, {\rm g} \sim 10^{-12} M_{\odot}$, however wave optics effects make it impossible to exclude $f_{\rm PBH} =1 $ (for a DF MF) for masses significantly smaller than this with optical microlensing observations \citep{Sugiyama:2019dgt}. New techniques will therefore be needed to probe the whole of the current asteroid mass window, see Sec.~\ref{subsec:asteroid}.

\subsection{Constraints on realistic extended mass functions}
\label{subsec:emf}

The constraints displayed in Fig.~\ref{fig_constraints} are all calculated assuming a DF MF. As discussed in Sec.~\ref{subsec:refinements} we don't expect PBHs formed from the collapse of large density perturbations to have a DF MF. Even if all PBHs form at the same time, from a sharp peak in the primordial power spectrum, critical collapse means that they will have a range of masses. For  extended mass functions the constraints are `smeared out'. For any given constraint the tightest value of the constraint (as a function of the central or peak mass) becomes weaker, however the constraint now applies (i.e.~$f_{\rm PBH} <1$) over a wider range of peak masses~\citep{Carr:2017jsz}. Therefore PBHs with realistic extended mass functions are more tightly constrained, in that $f_{\rm PBH} =1$ is excluded over a wider range of masses~\citep{Green:2016xgy,Carr:2017jsz}.

\section{Open questions}
\label{sec:openq}
In this section we discuss three of the key open questions in the field of PBH dark matter, namely how to probe asteroid mass PBHs observationally (Sec.~\ref{subsec:asteroid}), the probability distribution of density perturbations produced by ultra-slow-roll inflation (Sec.~\ref{subsec:USR}) and the clustering of PBHs on subgalactic scales (Sec.~\ref{subsec:clustering}).

\subsection{How to probe asteroid mass PBHs?}
\label{subsec:asteroid}

As discussed in Sec.~\ref{subsec:futureconstraints}, future observations will increase the range of PBH masses probed by the evaporation and microlensing constraints somewhat. However it will not be possible to probe all of the asteroid mass window, $10^{17} \, {\rm g} \lesssim M_{\rm PBH} \lesssim 10^{22} \, {\rm g}$, using these techniques. To do this other methods will be required. 

\begin{figure}
	\centering 
	\includegraphics[width=0.5\textwidth, angle=0]{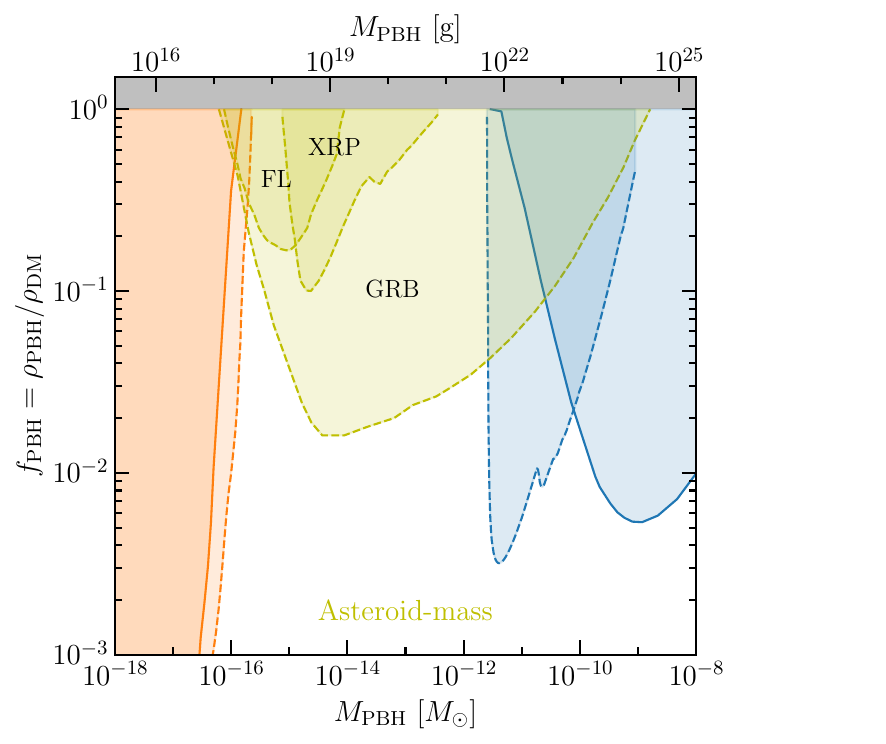}	
	\caption{Current and projected future (solid and dotted lines respectively) constraints on the fraction of DM in the form of asteroid mass PBHs, $f_{\rm PBHs}$, as a function of mass, $M_{\rm PBH}$, assuming all PBHs have the same mass. 
   The bounds shown are from evaporation (in orange), optical microlensing (blue) and other methods, involving lensing of X-ray pulsars~\citep{Bai:2018bej} and gamma ray bursts~\citep{Katz:2018zrn,Jung:2019fcs}, (yellow). The future evaporation and microlensing constraints are from \citet{Coogan:2021rez} and \citet{Sugiyama:2019dgt} respectively. See text for further discussion. Figure created using the PBHbounds code~\citep{PBHbounds}.} 
	\label{fig_asteroid}%
\end{figure}

Lensing observations using X-rays and gamma-rays can avoid the wave optics limitations which apply to optical microlensing observations. One possibility is femtolensing of gamma ray bursts (GRBs)~\citep{1992ApJ...386L...5G}, however in this case observations of small GRBs are required to limit the reduction in magnification from finite source size effects~\citep{Katz:2018zrn}. Other possibilities include GRB lensing parallax~\citep{Nemiroff:1995ak,Jung:2019fcs}, microlensing of X-ray pulsars~\citep{Bai:2018bej} and the effects of PBH encounters with stars (see e.g. \citet{Montero-Camacho:2019jte} and references therein and thereof). Figure~\ref{fig_asteroid} shows the current and projected future constraints for asteroid mass PBHs.

\subsection{What is the probability distribution of density perturbations produced by ultra-slow-roll inflation?}
\label{subsec:USR}

As discussed in Sec.~\ref{subsec:inflation}, in single-field inflation models to generate large, PBH forming density perturbations a period of ultra-slow-roll inflation (driven, for instance, by the slope of the potential becoming small) is required. In this case stochastic effects are important for the evolution of the scalar field, and the tail of the probability distribution of the size of the fluctuations could be gaussian rather than exponential (see e.g.~\citet{Pattison:2021oen,Figueroa:2020jkf,Tada:2021zzj} and references therein and thereof). This would have a significant effect on the abundance of PBHs formed and is the subject of extensive ongoing work.

On a different, but related, note there is currently ongoing discussion about whether large amplitude small scale perturbations lead to significant one-loop corrections to perturbations on CMB scales, and are hence excluded. See e.g.~\citet{Kristiano:2022maq,Firouzjahi:2023ahg,Fumagalli:2023hpa,Firouzjahi:2023bkt}.

\subsection{Clustering (on subgalactic scales)}
\label{subsec:clustering}

On galactic and larger scales PBH DM would be distributed in the same way as any other viable cold dark matter candidate. However their macroscopic nature may lead to enhanced clustering on subgalactic scales, which could have important consequences for the PBH binary merger rate (and potentially other constraints). As PBHs are discrete objects there are Poisson fluctuations in their distribution, and the primordial perturbations therefore have an additional isocurvature component~\citep{Afshordi:2003zb,Inman:2019wvr}. Consequently if PBHs make up a significant fraction of the DM, PBH clusters start forming shortly after matter-radiation equality~\citep{Afshordi:2003zb,Inman:2019wvr}. The evolution of these PBHs clusters (and the PBH binaries within them) to the present day is a challenging open problem~\citep{Jedamzik:2020ypm,Trashorras:2020mwn}.

If the DM is a mixture of PBHs and particle DM then the PBHs will accrete a halo of particle DM during matter domination, see e.g.~\citet{Mack:2006gz,Adamek:2019gns}. Consequently, PBHs and Weakly Interacting Massive Particles (WIMPs) both making up a significant fraction of the DM is essentially already excluded, as WIMP annihilation in the halos surrounding PBHs would produce a very large flux of gamma-rays~\citep{Lacki:2010zf}.


\section{Summary}
\label{sec:summary}

Primordial Black Holes can form in the early Universe, for instance from the collapse of large density perturbations during radiation domination. To produce an interesting number of PBHs, (i.e.~for them to make a non-negligible contribution to the present day energy density of the Universe) the amplitude of the primordial perturbations must be several orders of magnitude larger on some small scale than it is on cosmological scales. This can be achieved in some inflation models, e.g.~with a feature in the potential or multiple fields, however it is not generic.

There are numerous observational constraints on the abundance of PBHs from gravitational lensing, their evaporation, dynamical effects, accretion and other astrophysical processes. There have been significant developments in these constraints in recent years. New constraints have been proposed, and some old constraints have been found to be weaker than originally thought, or even shown to not be valid. Currently it appears that planetary, Solar and multi-Solar mass PBHs can't make up all of the dark matter, but lighter, asteroid mass, $M_{\rm PBH} \sim (10^{17}-10^{22}) \, {\rm g}$, PBHs could. Collectively the limits are tighter (i.e. $f_{\rm PBH} =1$ is allowed for a narrower range of masses) for realistic extended mass functions, than for the delta-function mass function which is usually assumed when calculating constraints. 

There are various important open problems in the field of PBH dark matter. These include observational probes of asteroid mass PBHs, the probability distribution of the density perturbations in ultra-slow-roll inflation models which generate large perturbations, and the present day clustering of PBHs on subgalactic scales and its effect on observational constraints/signals.  

\section*{Acknowledgements}

AMG is supported by a STFC Consolidated Grant [Grant No. ST/T000732/1].
For the purpose of open access, the author has applied a CC BY public copyright licence to any Author Accepted Manuscript version arising. \\

{\bf Data Availability Statement:} This review paper has no associated data.




\bibliographystyle{elsarticle-harv} 
\bibliography{PBHs}






\end{document}